# Quantifying Defects and Finite Size Effects in Graphene Oxide Models

Sownyak Mondal, Soumya Ghosh*

*Tata Institute of Fundamental Research Hyderabad, Hyderabad, 500046, Telangana, India*
*email: soumya.ghosh@tifrh.res.in*

**Abstract**: Oxidation of graphite and subsequent exfoliation leads to single layer graphene oxide (GO). GO has found many applications across diverse fields including medicinal chemistry, catalysis as well as a precursor for graphene. One of the key structural features of GO is the presence of different kinds of defects. Molecular dynamics simulations with ReaxFF force fields have been widely used to model realistic representations of GO that include defects of various types. In these simulations, one can vary the extent and distribution of the defects by changing the initial O/C ratio. It is therefore very important to employ a proper measure of the defect density. Traditionally, the total number of non-graphitic carbon atoms have been employed to quantify the amount of defects. Our simulations suggest that this parameter may not be a good measure at low defect densities. Herein, we introduce a hitherto unexplored metric, relative area of the defects, to gauge the defect density. We show that this metric has desirable properties at both low and high defect densities. Additionally, we investigate the changes in the defect distribution and mechanical properties upon varying the size of the simulation cell.

## Introduction

The single-layered oxidized form of graphene, graphene oxide (GO), has numerous applications.[1,2] Different GO composites have been employed in tissue engineering and drug delivery.[3] GO supported transition metal complexes can catalyze several important organic and inorganic transformations under ambient and electrochemical conditions.[4,5] Solubility in water and ease of synthesis of GO has been exploited for large scale graphene synthesis from graphite.[6] Several other applications, e.g. as a membrane, relies on controlling the physical and chemical properties of GO by varying the extent of oxidation and the distribution of different oxygen species on the surface.[7–9] Similarly, the electronic properties (e.g. band gap) of GO can be tuned based on the degree of oxidation making it an attractive material for optoelectronic devices.[10–12]

Extensive simulations have been carried out on GO to understand its chemical and physical properties. One of the key features of GO structures is the presence of different kinds of non-hexagonal regions (defects), whose density is a function of the extent of oxidation.[13,14] ReaxFF[15,16] simulations can provide access to different types of defects in GO and its derivatives,[17] and hence, ReaxFF based molecular dynamics (MD) simulations were performed to understand several physical processes involving GO, e.g. friction behavior of monolayer GO confined between amorphous iron oxide substrates, [ref.] fatigue failure in GO,[18] impact of GO intercalation on the tensile strength of cement hydration products,[19], the thermal conductivity of functionalized GO,[20] and electric double layer capacitance of GO.[21] Moreover, the ability of ReaxFF simulations to capture bond rearrangements were exploited to understand mechanisms of several chemical processes, e.g. interaction of GO with water,[22] crystal growth of nanoconfined intercalated Cu in GO,[23] catalytic depolymerization of polyethylene using GO supported Pt nanoparticles,[24] nitrogen doping in GO,[25] separation of metal ions and liquid solubites (alcohol) from water using membranes of GO or its composites,[26] electrochemical hydrogen production activity of BCN co-doped GO,[27] microwave reduction of GO,[28] and cell capture by functionalized GO.[29]

Quantification of the defect densities has been done in various ways. In the seminal work by Lucchese et al, careful analysis of the STM images of the graphene sheets obtained after bombardment of different concentrations of $Ar^+$ ions provided an approximate estimate of the distance between the defects.[30] The trend in the intensity ratio of the defect and graphitic peaks in Raman spectra of graphene has been employed to extract defect densities in GO.[31] Alternatively, Perozzi et al coupled optical microscopy with X-ray photoemission spectroscopy to determine the fraction of $sp^2$ hybridized carbon atoms.[32] High resolution TEM images have also been used to track the graphene like regions and different kinds of defect polygons in reduced graphene oxides.[33] The estimation of the defect densities obtained from model simulations is mostly based on the count/fraction of the $sp^2$ carbons.[34] Alternatively, defect pore size has been estimated by determining the diameter of the largest possible inscribed sphere.[35]

Given that the properties of GO are quite dependent on the local heterogeneity of the structures, we ask the question whether the defect distribution changes with the simulation cell size. Additionally, we also check the mechanical response of the systems with different dimensions at various defect densities to figure out whether there is any aspect that is system size independent. Finally, we introduce a new metric to gauge the defect densities and compare the results with the percentage of non-graphitic carbon atoms.

**Computational details:**

The current study employs molecular dynamics (MD) simulations with the ReaxFF force field parameters.[36] The orthorhombic unit cell dimensions of the underlying graphene structure is b = 4.26 Å, a = 2.46 Å.This unit cell is repeated multiple times in the X and Y directions to obtain the different models that are employed in the present study: G1 (12,7) with 360 C atoms, G2 (17, 10) with 680 C atoms, and G3 (20, 12) with 960 C atoms, respectively. Depending on the desired oxygen to carbon ratio, the hydroxyl and epoxy groups were placed randomly on both sides of the graphene sheets in a 1:1 ratio with the constraint that no two oxygen atoms occupied the neighboring carbons. The Atomic Simulation Environment [37] package was employed to build the neighbor list. In order to generate the final structures with defects, we considered the following protocol: (1) The temperature of the system is increased to 1500K within 250 ps. (2) The system was then annealed for 200 ps at 1500K and subsequently, (3) it was quenched to 300K and equilibrated for 5 ps. All the simulations were performed using Large-scale Atomic/Molecular Massively Parallel Simulator (LAMMPS).[38] Berendsen thermostat was employed to modulate the temperature in all three NVT ensembles.

We investigated the stress-strain response of all the systems using the athermal quasistatic simulation (AQS) method.[39] This protocol is followed to probe the shear-induced response of the system in the absence of any kind of thermal fluctuations. The finite temperature graphene oxide structure (taken from the previous process) has been minimized to an energy minimum at zero temperature using the conjugate gradient algorithm. Then two iterative processes have been followed. First, the system is deformed by a slight amount by applying a simple shear transformation (affine) to each particle. After every deformation step, the potential energy of the deformed system is minimized using the conjugate gradient algorithm. The Lees-Edwards boundary condition has been used in both steps.

In order to analyze the defect polygons, one would need to track them first. In principle, one can track them by employing algorithms related to non-directed cyclic graphs if one takes care of the boundary conditions carefully.[40] In this work, we isolate the polygons using an algorithm based on minimum dihedral criterion as one moves from one center to its neighbor. Once a polygon (defect or non-defect) has been isolated, we compute its area. This exercise is non-trivial because the polygons are not planar. We employ the algorithm developed by Zhou et al. for triangulation of non-planar 3D polygons that minimizes the sum of areas of all the triangles. This complex triangulation scheme starts by determining a pool of candidates using the set of triangle facets in the Delaunay tetrahedralization of the polygon vertices and then divides the total polygon into domains whose endpoints share an edge in the input triangle set. This strategy is repeated recursively until the optimal triangulation with the prescribed criterion is satisfied.[40] We employed the python library trimesh to compute the discrete Gaussian curvature (K) of the triangulated defect polygons.[41] The radii of all the vertices are set to the covalent radius of the carbon atom (0.72 Å).

**Results and Discussion**

The structure of GO is riddled with different kinds of defects. In this work, the defects are categorized into two groups based on the number of edges that constitute the defects: type-1, where the number of edges is between 5-8 (barring 6), and type-2, where this number is greater than 8 (Figure 1).

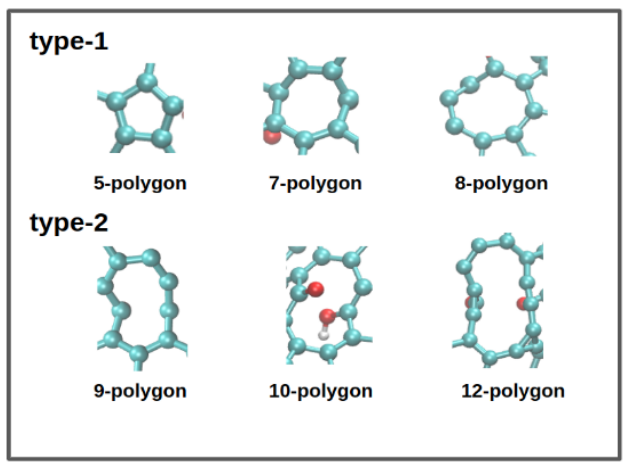

**Figure 1**. Different kinds of defects are categorized into two types based on size: type-1 (moderate size: 5 + 7 + 8 membered polygons) and type-2 (large size: polygons larger than 8 sides)

The distribution of type-1 and type-2 defects as a function of initial O/C %, is computed for 3 different sizes of the graphene sheets – G1, G2, and G3 as mentioned in the computational details section. The percentage of the two types of defects for four different initial O/C ratios is shown in Figure 2. While there is hardly any monotonic feature, some qualitative trends are apparent. G1 seems to favor Stone-Wales defects whereas G2 shows preference for vacancy defects at low to moderate defect densities but percentage of Stone-Wales starts to dominate at higher defect densities. G3, on the other hand, exhibits defect distribution akin to G1 at both low and high defect densities but seems to prefer vacancy defects, similar to G2, at moderate defect densities.

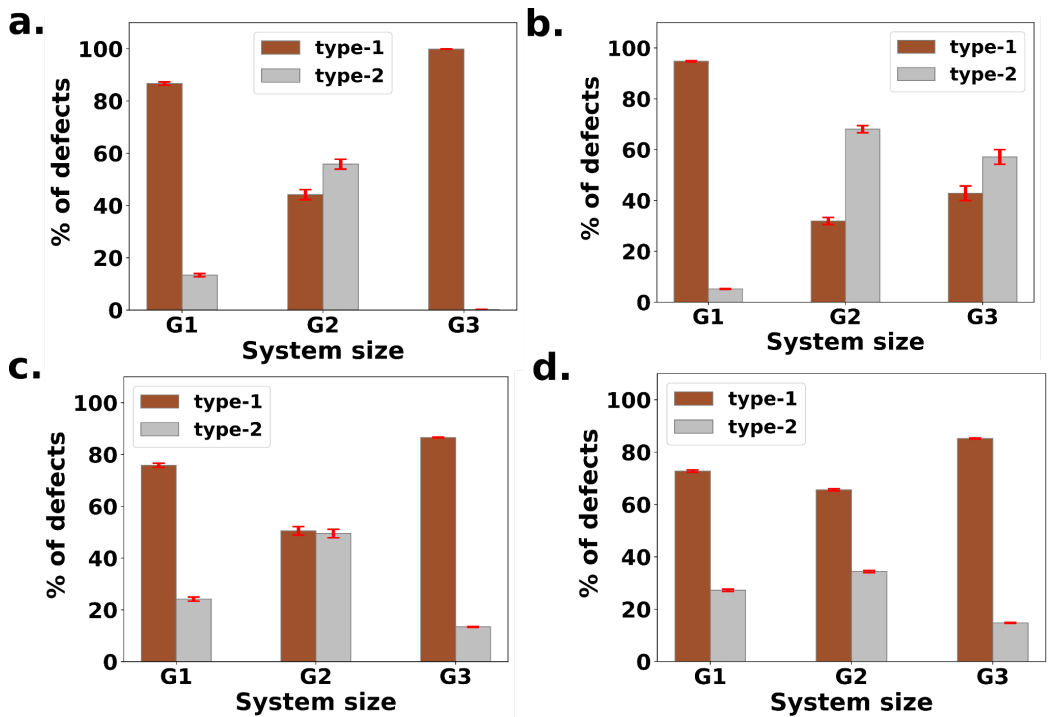

**Figure 2**. Distribution of type-1 and type-2 defects for different initial O/C ratios (a: 5%, b: 10%; c: 15% and d: 20%) for simulation cells with 336 (G1), 680 (G2), and 960 (G3) carbon atoms

The lack of a clear trend in the distribution of defects at low to moderate defect densities (measured by initial O/C ratio) indicates that the defect distribution might be dependent on the initial distribution of the oxygen containing groups. In order to test this hypothesis, we generated the initial oxygen distribution in G2 (5%) by taking the oxygen distribution from G1 (10%) and found that the type-2 defect in G2 has disappeared (Figure 3).[1] Similarly, starting with an initial oxygen distribution in G3 (5%) akin to oxygen distribution in G2 (10%) induces a type-2 defect in G3 (5%).

---

[1] *The oxygen distribution for the larger size cell is obtained from the smaller size cell at higher O/C ratio in order to avoid unrealistic increase in oxygen density at the periphery.*

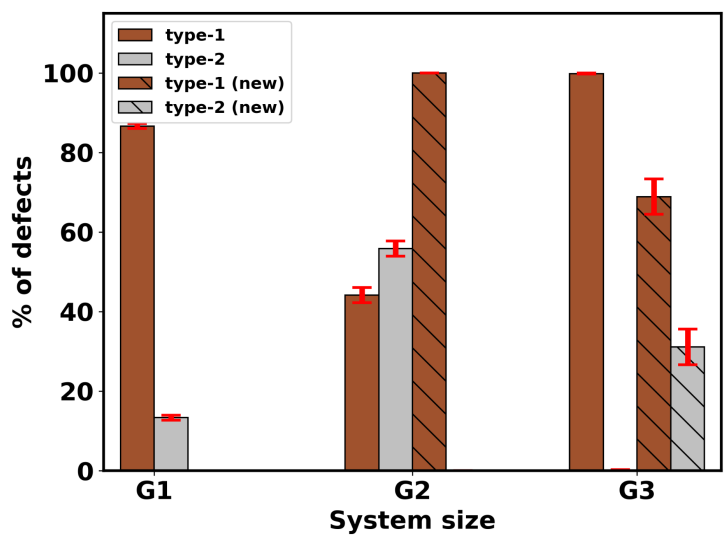

**Figure 3**. Variation in the distribution of defects due to different initial distributions of the O containing groups on the carbon backbone. The initial O/C ratio is maintained at 5%.

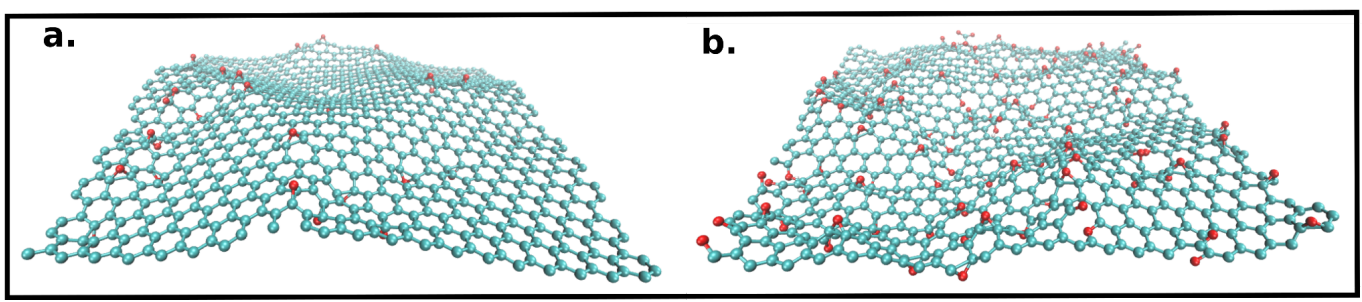

**Figure 4.** G3 structure at (a) low (5%) and (b) high (20%) defect densities

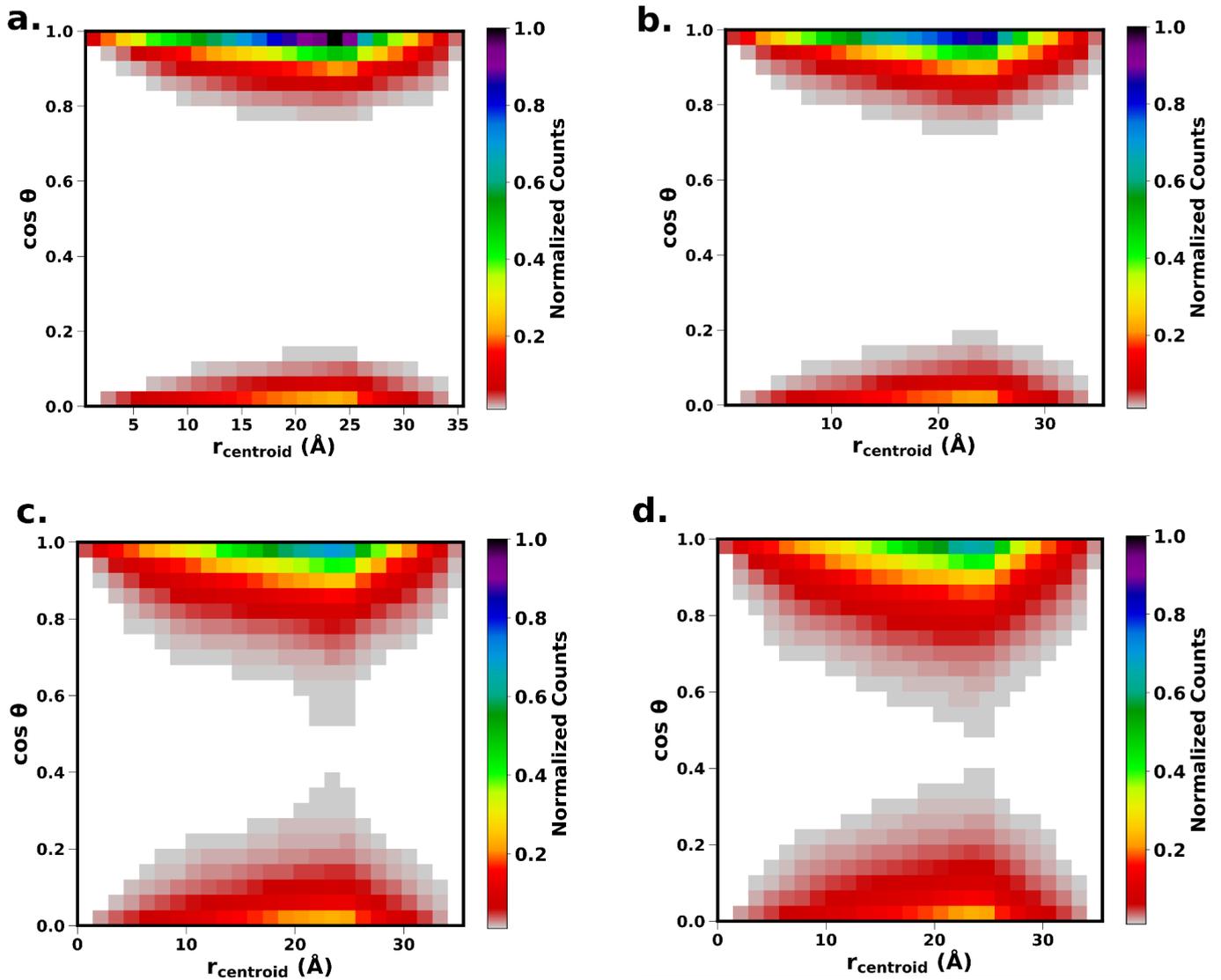

**Figure 5**. The local deformations in the G3 structure is quantified by correlating the normal vectors on the triangulated surface as a function of the separation between the centroids of the triangles for the following defect densities: (a) 5% (b) 10% c) 15% (d) 20%

With the increase in defect density the surface becomes more irregular (Figure 4). In order to provide a quantitative estimation of the extent of deformation, we first triangulated all the polygons on the surface and then computed the orientational correlation between the normals on the triangles as a function of separation between the centroids of the triangles (Figure 5). The heat maps clearly show that the normals become more uncorrelated with the increase in defect density. The drastic change in the local structure of the surface with the increase in defect density may imply that the 3D structures of defect polygons will also vary with the defect density. In order to address this aspect, we explored whether the defect polygons have unique topologies, measured by gaussian curvatures (K), along with distinct oxygen densities across different system sizes and defect densities. The oxygen densities are reported as the number of oxygen atoms present per unit area of the defect structures, which are considered to be 3D polygons as described in the computational details. The complete data is presented as a heatmap in Figure 6. We find that 5-membered and 7-membered polygons have similar gaussian curvatures over a wide range of oxygen densities with the maxima near 0.14 Å$^{-2}$. Similarly, 8-, 9/10-, 11-, 12/13- and 14/15-membered polygons have unique gaussian curvatures around 2.425, 2.315,

2.278, 2.242, and 2.205, respectively. The corresponding variation in oxygen densities can be broad or narrow.

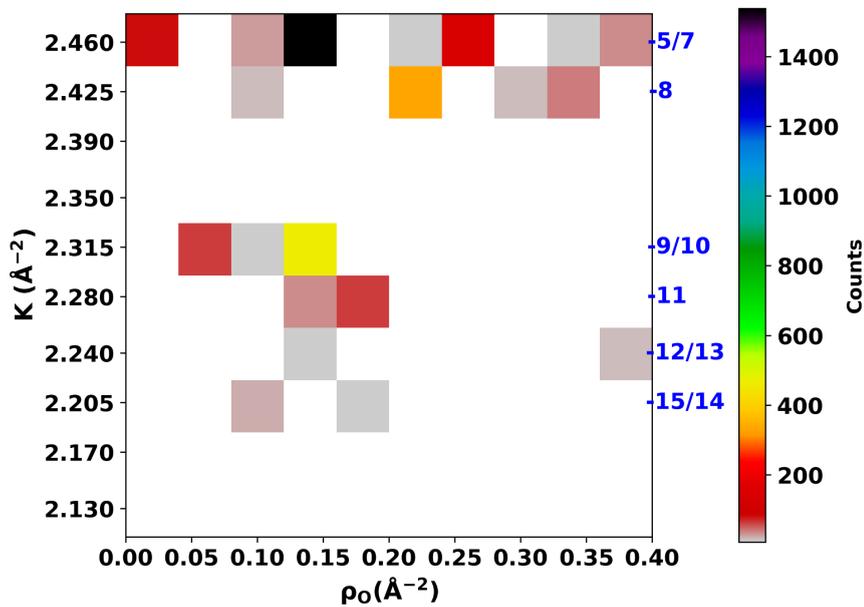

**Figure 6**. The gaussian curvatures of different sized polygons (indicated by blue labels on the right side of the plot) are mapped along with the areal density of the oxygen atoms in the defects.

The variation in the distribution of the different types of defects with system size begs the question whether the mechanical properties are system size dependent or not. The response of the GO structures to external shear is often taken as a gauge to estimate their mechanical strengths.[41,42] In Figure 7(a,b) the stress-strain response is plotted for GOs of different sizes with initial O/C ratio of 5 % and 15 % respectively. In both the cases, the yield strain remains approximately the same irrespective of the system size but the upper stress point is sensitive to the size. It can also be seen that several plastic events precede the yield point. The number of plastic events increases with the increase in O/C ratio. The shear strain at which these plastic events occur is also sensitive to the system size. A comparison of shear modulus for G1, G2, and G3 for different O/C ratios is shown in Figure 7c.

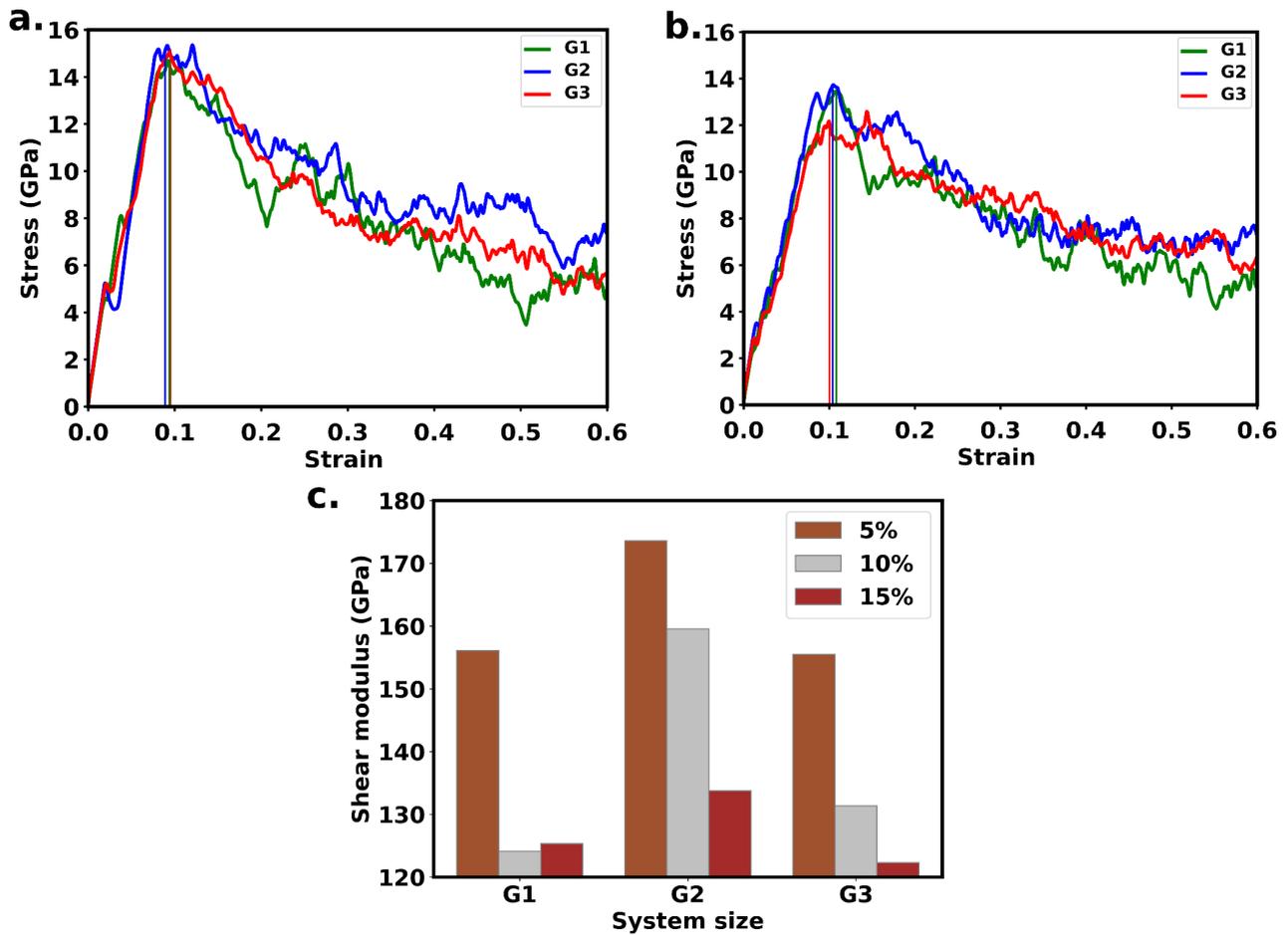

**Figure 7**. (a), (b) Stress-strain response of the equilibrated structures with initial O/C ratio of 5 % and 15 %, respectively. (c) Comparison of the shear modulus for different system sizes and different initial O/C ratios.

It is clear from the above results that in order to properly track the defect density one would need a parameter that is sensitive to the local deformations in the structures. First we examine how the widely employed parameter, the total percentage of the non-graphitic carbon atoms, fare with the change in the initial O/C ratio. As expected, this parameter increases, more or less monotonically, with the increase in the O/C ratio as depicted by the solid black line in Figure 8a for the G3 system. At low defect density corresponding to a small O/C ratio, there is, however, an appreciable fraction of non-sp2 carbon atoms that are not part of the defects as shown by the dotted line in Figure 8a. A couple of representative examples are provided in Figure 8b. Hence, it seems that one should track the percentage of non-graphitic carbon atoms that are *exclusive to the defects*, dashed line in Figure 8a, in order to better track the defects at low densities. On the other hand, at large defect densities, there are only a few defect-free regions and even a small percentage of non-graphitic carbon atoms that are not part of any defect can affect the local structural morphology of the system. Therefore, it might not be adequate to measure just the fraction of the defect exclusive non-graphitic carbon atoms for tracking the defect densities at large defect densities. In short, ideally one would want a parameter that will follow the defect exclusive non-graphitic carbon atoms at low defect densities but would increase more sharply at high defect densities.

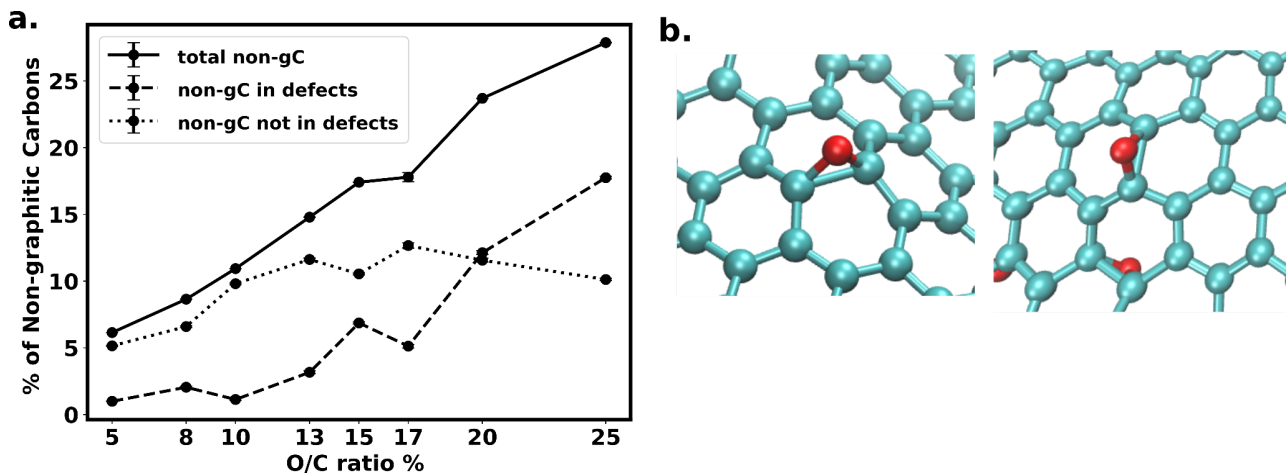

**Figure 8**. (a) Change in the percentages of the non-graphitic carbon atoms as a function of initial O/C ratio. At low O/C ratios the percentage of the total non-graphitic carbon atoms is dominated by the percentage of non-graphitic carbon atoms that are not part of any defect (b) Examples of non-graphitic carbon atoms, one each from structures obtained with initial O/C ratios of 5% and 20% respectively, that are not part of the defects but may contribute to the local structural deformations.

In this study, we introduce an alternative parameter that tracks the defects exclusively but has some additional desired features. As mentioned in the computational details, we consider the defect structures as 3D polygons and triangulate them in such a way that the total area of all the triangles in a defect is minimized [43]. An example of such a triangulated defect is shown in Figure 9a. The versatility of this parameter, in comparison to the traditional parameter, is demonstrated in the following 3 scenarios: (1) by analyzing the equilibrated structures with different defect densities obtained from different initial O/C ratios (2) during the formation of defects starting from the initial pristine planar geometry with randomly placed hydroxyl and epoxy groups following the protocol described in the computational details (3) during the application of external shear on the equilibrated structures with different defect densities. All the analyses are performed with G3.

The variation in the defect area with O/C ratio is compared with the percentages of the total and defect exclusive non-graphitic carbon atoms in Figure 9b. The defect area at low defect densities follow the defect exclusive non-graphitic carbon atoms but increases more rapidly at higher defect densities. As discussed earlier, with the increase in the defect density, the non-graphitic carbon atoms that are not part of the defects can cause appreciable distortion to the structure. Hence, the defect area, which is sensitive to the local distortions of the structure, increases more sharply than the non-graphitic carbon atoms that are exclusive to the defects and seems to touch the values corresponding to the total non-graphitic carbon atoms at very high defect densities.

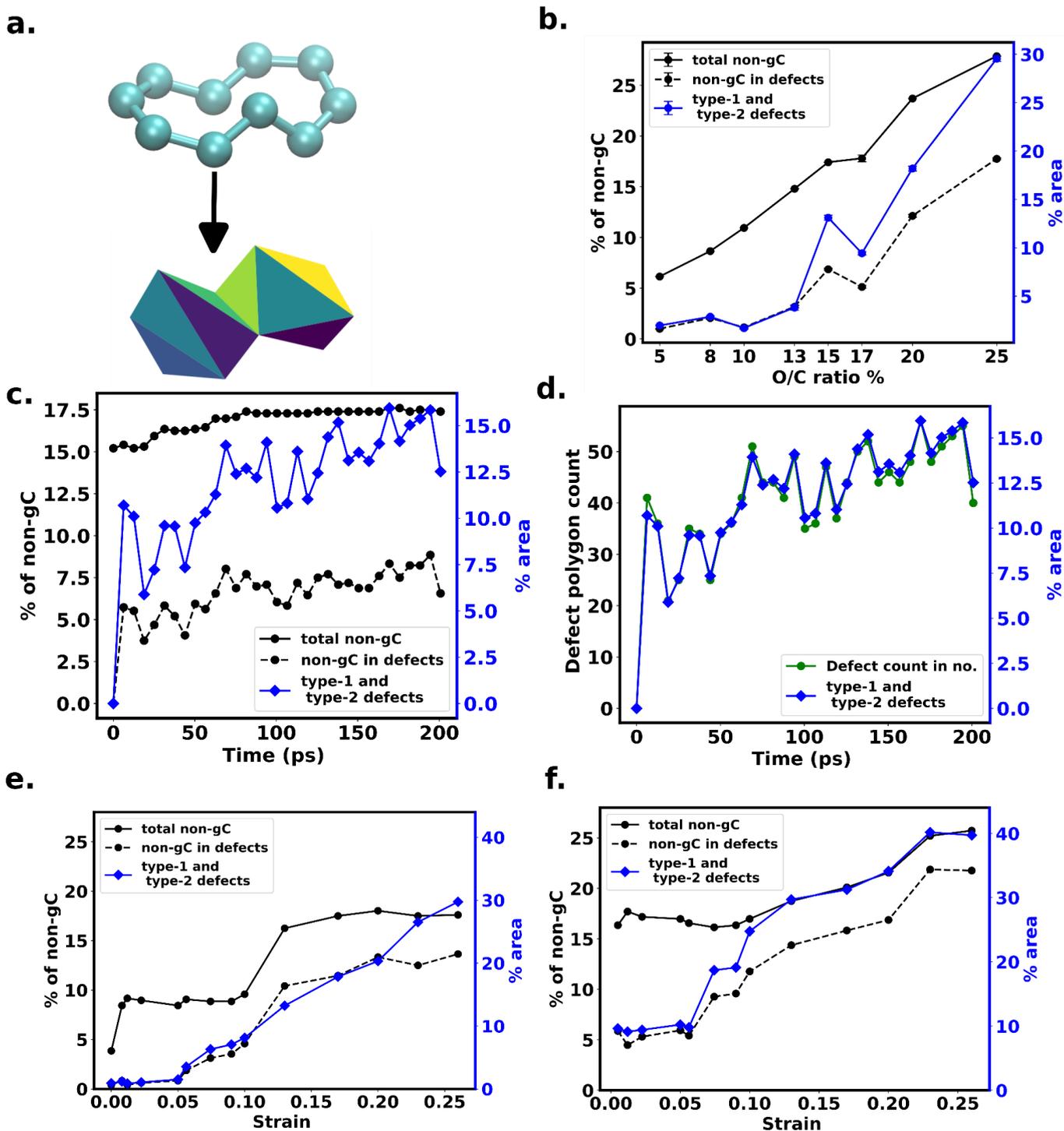

**Figure 9**. (a) An example of a triangulated defect structure (b) Area of defect polygons in 3D is compared to the number of non-graphitic carbon atoms that are *exclusive to the defects* and the total number of non-graphitic carbon atoms as a function of the initial O/C ratio (c) Evolution of the three parameters used to characterize the defects during the MD run starting from the initial pristine structure (d) Evolution of the number of defect polygons vs the defect area during the MD run (e) (f) Change in the % of the non-graphitic carbon atoms (total and defect exclusive) and percentage of the defect area as a function of applied shear strain for 5 % and 15% initial O/C ratio, respectively

Following the protocol described in the computational details, the defect structures are formed from the pristine oxidized graphene in a span of ~ 200 ps. The evolution of the three parameters (the percentages of the total and defect exclusive non-graphitic carbon atoms and the percentage of the defect area) used to characterize the defect densities is shown in Figure 9c. In this example, we have chosen the initial O/C ratio to be 15% on the G3 system. Most of the defects (~ 42 in number) are formed within the first few picoseconds but the relative number of the different kinds of defects keep changing. While the percentage of the total number of non-graphitic carbon atoms remains more or less constant, the change in the number of defect exclusive non-graphitic carbon atoms is more pronounced. The difference between the percentages of the total and defect exclusive non-graphitic carbon atoms is significant, as pointed out earlier, owing to the presence of a substantial number of non-graphitic carbon atoms that are not part of the defects. The overall change in the percentage of the total defect area over 200 ps is also small but the fluctuations are much more significant. The fluctuation in defect area closely follows the fluctuation in the total number of defect polygons (Figure 9d). Hence, in the case of the evolution of defects from the initial pristine oxidized structure the scaled plot for the total defect area increases faster than the percentage of the defect exclusive non-graphitic carbon atoms, similar to the trend observed in Figure 9b.

The change in the percentages of the total and defect exclusive non-graphitic carbon atoms during the application of shear strain for 5 and 15 % initial O/C ratios are compared with the change in the percentage of the defect area in Figure 9e,f. In both the cases, the percentage of the total non-graphitic carbon atoms remains constant till the yield strain (~ 0.1) and then increases slowly. On the other hand, the percentage of the defect exclusive non-graphitic carbon atoms seems to grow monotonically with the application of strain at both 5% and 15% O/C ratios. Further analysis of the structures generated upon application of the shear strain sheds light on the difference in the trend of the total and defect exclusive non-graphitic carbon atoms prior to the yield strain. When the system is subjected to external shear new defects are formed encompassing the non-graphitic carbon atoms that were initially not part of any defect (Figure 10). This process continues till the yield strain. Beyond this point, generation of new defects necessitates the formation of new non-graphitic carbon sites and hence, the two graphs run parallelly.

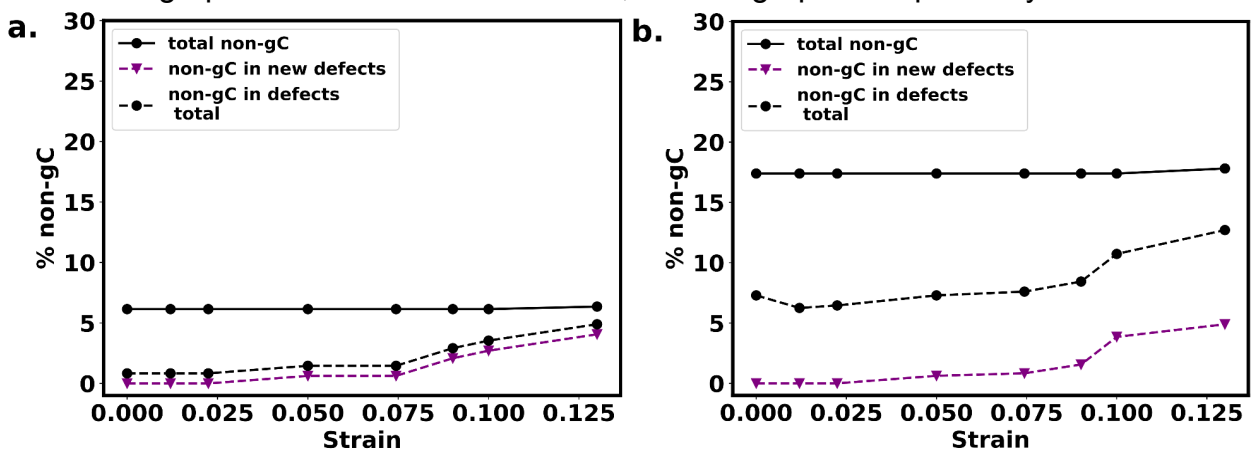

**Figure 10**. Change in the percentage of the non-graphitic carbon atoms that were initially not part of any defect for (a) 5 % O/C ratio (b) 15% O/C ratio

The defect area seems to follow the defect exclusive non-graphitic carbon atoms till the yield strain. Beyond the yield strain the defect area grows more rapidly than the defect exclusive non-graphitic carbon atoms (Figure 9e,f). Further analysis of the structures generated due to application of the external shear reveals that beyond the yield strain there is an increase in the number of irregular shaped large defects (type-2), sometimes by merging the existing smaller defects (Figure 11), a feature that is also nicely captured by the sharp increase in the resulting defect area.

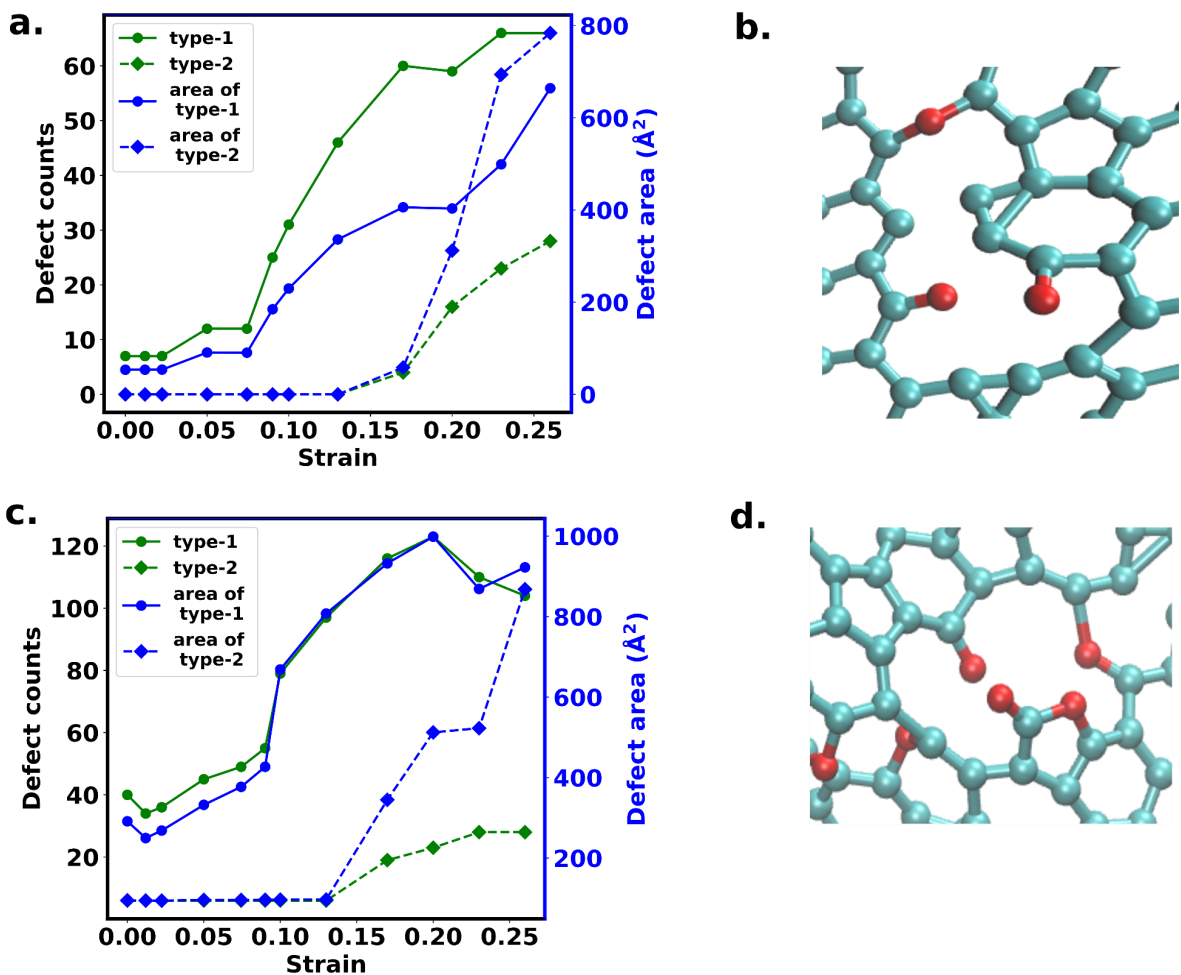

**Figure 11**. The change in the count of type-1 and type-2 defects and their corresponding area is plotted against the applied strain for (a) 5% (c) 15% O/C ratios. Representative examples of a large sized defect for the two percentages are provided in (b) and (d).

## Conclusions

In this study we showed that the relative abundance of the various kinds of defect polygons that are found in the simulated GO structures vary extensively with the system size and initial distribution of oxygen atoms on the pristine graphene structure, due mainly to the local deformations of the structures. On the other hand, the strain at which the system reaches the upper stress limit upon application of external shear remains approximately invariant to the system size. The upper stress limit itself and consequently the shear modulus, however, depends on the size of the simulation cell. Additionally, we introduced a new metric, the minimum area enclosed by the neighboring defect sites constituting the non-hexagonal 3D polygons, to gauge the defect densities. In general, we found that the percentage of the total non-graphitic carbon atoms might be able to track the defects adequately in the high defect density regime whereas the percentage of the defect exclusive non-graphitic carbon atoms is a better metric at low defect densities. In contrast, the defect area, which is sensitive to the local structural deformations, is able to track the defects well across all the regimes.

**Acknowledgments**

SM and SG would like to thank Prof. T N Narayanan, Manoj Adhikari, and Sumit Bawari for insightful discussions. SG would also like to acknowledge the TIFRH HPC facility for computational resources and SERB/SRG grant for support.